\documentclass[epj]{svjour}
\usepackage{graphicx}
\usepackage{xspace}
\usepackage[tight]{subfigure}

\newcommand{\pT}{\ensuremath{p_T}\xspace}
\newcommand{\vTwo}{\ensuremath{v_2}\xspace}
\newcommand{\vTwoRaw}{\ensuremath{v_2^{raw}}\xspace}
\newcommand{\vTwoCorr}{\ensuremath{v_2^{corr}}\xspace}
\newcommand{\piz}{\ensuremath{\pi^0}\xspace}
\newcommand{\RAA}{\ensuremath{R_{AA}}\xspace}
\newcommand{\photon}{\ensuremath{\gamma}\xspace}
\newcommand{\meanNpart}{\ensuremath{\langle N_{part}\rangle}\xspace}
\newcommand{\dphi}{\ensuremath{\Delta\phi}\xspace}
\newcommand{\leff}{\ensuremath{\rho~L~dL}\xspace}
\makeatletter
\newcommand*{\SubLabels}[1]{%
  \label{fig:#1}%
  \begingroup
    \protected@edef\@currentlabel{%
      \csname thesub\@captype\endcsname
    }%
    \label{subfig:#1}%
  \endgroup
}
\makeatother

\begin{document}
\title{High-\pT \piz Production with Respect to the Reaction Plane Using the
PHENIX Detector at RHIC}
\author{David L. Winter\thanks{\emph{email:}
winter@nevis.columbia.edu} for the PHENIX Collaboration
}                     
%
%
\institute{Columbia University, NY, NY 10027, USA}
\date{Received: date / Revised version: date}
%
\abstract{
The origin of the azimuthal anisotropy in particle yields at high \pT
($\pT>5$~GeV/c) in RHIC collisions remains an intriguing puzzle.
Traditional flow and parton energy loss models have failed to
completely explain the large \vTwo observed at high \pT.  Measurement of
this parameter at high \pT will help to gain an understanding of the
interplay between flow, recombination and energy loss, and the role
they play in the transition from soft to hard physics.  Neutral mesons
measured in the PHENIX experiment provide an ideal observable for
such studies.  We present recent measurements of \piz yields with
respect to the reaction plane, and discuss the impact current models
have on our understanding of these mechanisms.
} 
\authorrunning{D. L. Winter for the PHENIX Collaboration}
\titlerunning{High-\pT \piz Production with Respect to the Reaction Plane}
\maketitle
\section{Introduction}
\label{intro}

Two of the greatest mysteries that have arisen from the RHIC physics program
are the source of the apparent flatness of the high \pT ($>5$~GeV/c)
suppression of \RAA~\cite{ref:tadaaki_pizRaa} and the source of non-zero \vTwo
at high \pT~\cite{ref:highptv2}.  The existence of intermediate to high \pT
\vTwo was suggested early in the RHIC program~\cite{ref:gvw}, and has been the
subject of many theoretical treatments (see~\cite{ref:shuryak,ref:dfj} for
some additional examples).  Traditional flow and parton energy loss pictures
have failed to describe the magnitude of this anisotropy.  Measurement of the
azimuthal asymmetry $v_2$ at high $p_T$ will shed light on the contributions
from flow, recombination, and energy loss, as well as the transition from soft
to hard production mechanisms.

\section{Measuring \vTwo and \piz yields in PHENIX}
\label{sec:1}

The orientation of the reaction plane is measured event-by-event using the set
of two PHENIX Beam-Beam Counters (BBCs), which reside at the region
$3<|\eta|<4$.  Each detector is an array of 64 hexagonal, close-packed quartz
Cherenkov counters, located 150 cm from the interaction point.  The charge
measured by each counter is proportional (on average) to the multiplicity of
particles hitting it.  The reaction plane angle $\Psi_{RP}$ is determined from
the value of $\langle \cos 2\phi \rangle$.  Because the two BBCs provide
independent measurements of $\Psi_{RP}$, we can estimate the resolution of the
combined measurement via standard techniques~\cite{ref:resolution}.

For measuring photons and \piz{}s, we use the Electomagnetic Calorimeter
(EMCal)~\cite{ref:emcal}. Candidate clusters are required to pass \photon
identification cuts, and $m_{inv}$ distributions are formed from pairs of
these clusters.  The resulting yields are binned in angle with respect to the
reaction plane ($\Delta\phi = \phi - \Psi_{RP}$).  A similarly binned mixed
event background is then subtracted. The counts in the remaining peak centered
on the $\pi$ mass are integrated in a $\pm 2\sigma$ window (where $\sigma$ is
the width of a Gaussian fit to the peak).  Six bins in $\Delta\phi$ are used in
the interval $[0-\pi/2]$.

To measure \vTwo, we fit the raw (uncorrected) $\Delta\phi$
distribution $Y(\Delta\phi)$ as 
\begin{equation}
Y_{raw}(\Delta\phi) \propto 1 + 2\vTwoRaw\cos(2\Delta\phi).
\end{equation}
The resulting \vTwo parameter needs to be corrected for
the reaction plane measurement resolution, hence the designation \vTwoRaw.
The resolution $\sigma_{RP}$ is determined for each centrality bin, and leads
to the corrected value $\vTwoCorr = \vTwoRaw/\sigma_{RP}$.  The yields as a
function of $\Delta\phi$ can then be corrected with a factor
\begin{equation}
Y(\Delta\phi) = Y_{raw}(\Delta\phi) \times
\frac{1+2\vTwoCorr\cos2\Delta\phi}{1+2\vTwoRaw\cos2\Delta\phi}.
\end{equation}

\section{Results and Discussion}


\begin{figure}[t]
\begin{center}
\includegraphics[width=0.45\textwidth]{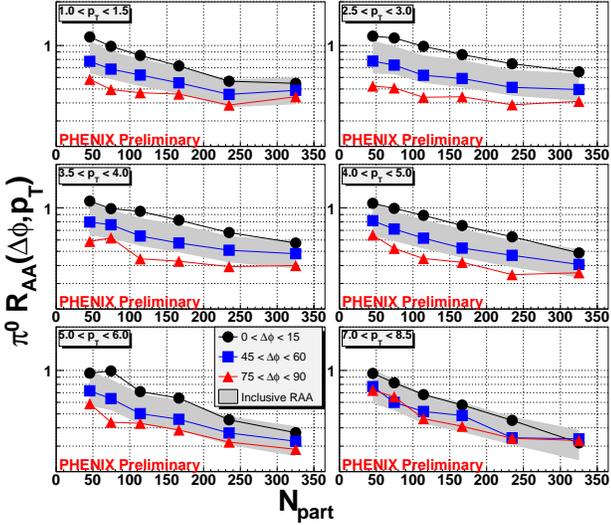}
\end{center}
\caption{$\RAA(\pT)$ as a function of \meanNpart for Au+Au 200 GeV/$c$
  collisions; Each curve corresponds to a different \dphi bin and the
  panels are for different \pT bins.  The grey bands indicate the
  systematic error due to the reaction plane resolution correction.}
\label{fig:raa_npart}
\end{figure}

\begin{figure}[t]
\begin{center}
\includegraphics[width=0.45\textwidth]{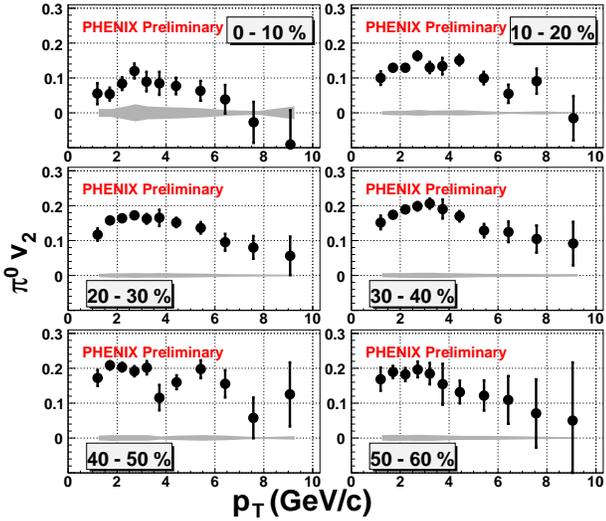}
\end{center}
\caption{\piz $\vTwo(\pT)$ for Au+Au 200 GeV/$c$ collisions, with each
  panel corresponding to a centrality bin.  The grey bands indicate
  the systematic error due to the reaction plane resolution
  correction.}
\label{fig:piz_v2}
\end{figure}

To obtain $\RAA(\Delta\phi)$, we exploit the fact that the ratio of the yield
at a given $\Delta\phi$ to the inclusive yield is equivalent to the ratio of
the angle-dependent \RAA to the inclusive \RAA.  Thus multiplying these
relative yields by an inclusive measured $\RAA$, we have:
\begin{equation}
\RAA(\Delta\phi) = Y(\Delta\phi) / Y \times \RAA
\end{equation}

The $\RAA(\Delta\phi,\pT)$ as a function of \meanNpart is shown in
Figure~\ref{fig:raa_npart}.  We note that there appears to be a
slightly different centrality dependence, according which \dphi bin in
which the \RAA is being measured.  This feature is emphasized by
plotting the data on a semi-log scale, showing that the \RAA behaves
differently in different \dphi bins.

The resulting \piz \vTwo is shown in Figure~\ref{fig:piz_v2}.  For the first
time we observe \vTwo up to 10 GeV/c.  While the value of the \vTwo decreases
beyond intermediate \pT, it nonetheless shows a substantial and
perhaps constant value out to the highest measured transverse momenta.

\begin{figure*}[t]
\begin{center}
\includegraphics*[width=\textwidth]{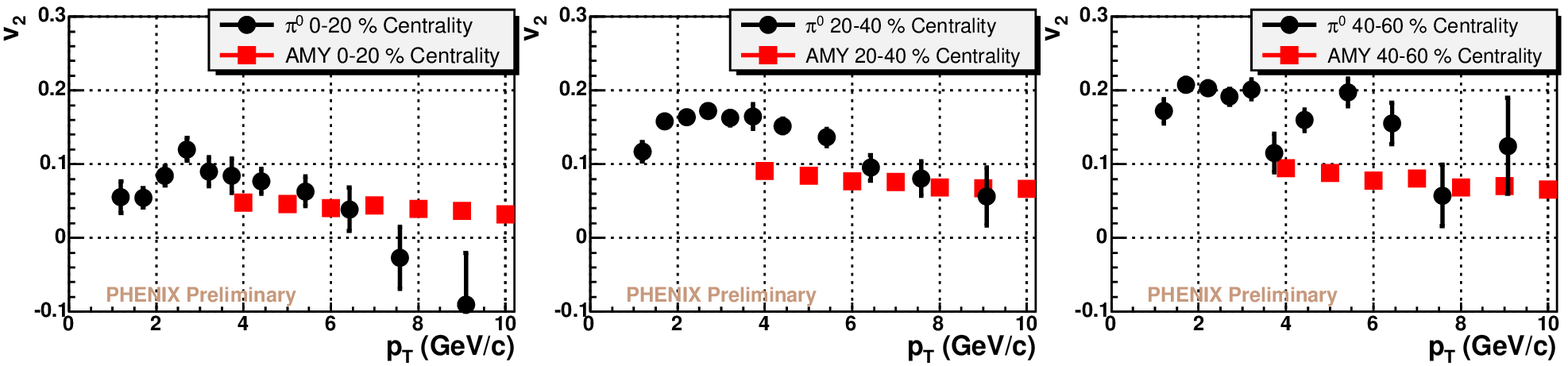}
\includegraphics*[height=4cm]{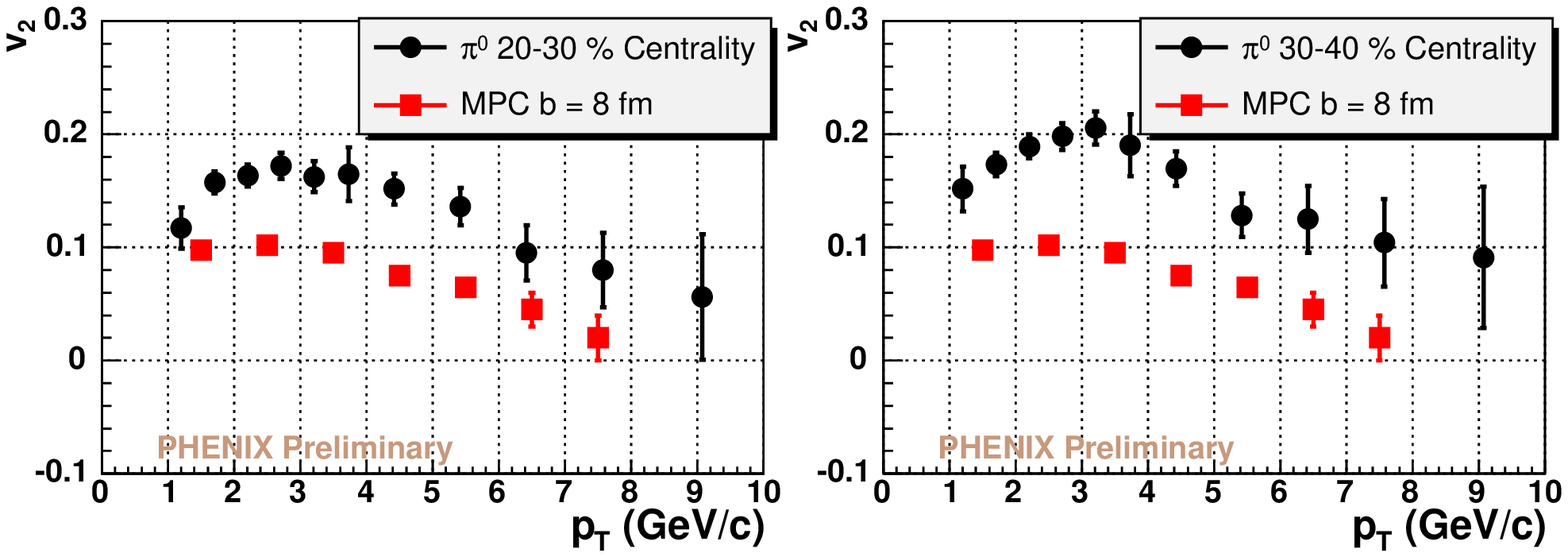}
\end{center}
\caption{Comparison of \piz \vTwo with models. The top three panels show the
  AMY calculation with data for three centralities~\cite{ref:turbide}.  The
  bottom two panels compare two centralities with the $b=8$~fm calculation of
  the MPC model~\cite{ref:mpc}.}
\label{fig:modelCompare}
\end{figure*}

To gain insight into the \vTwo mechanisms at work at high \pT, we turn to
models.  We compare the \piz \vTwo to two models, a calculation done by
Turbide et al.~\cite{ref:turbide} (using an Arnold-Moore-Yaffe (AMY)
formalism~\cite{ref:amy}) and the Molnar Parton Cascade (MPC)
model~\cite{ref:mpc}.  Figure~\ref{fig:modelCompare} shows calculations from
these models, plotted alongside data for similar centralities.  The AMY
calculation contains energy loss mechanisms only, and we see that the data
appear to decrease to a value at high \pT that is consistent with this model;
the level of agreement is most striking in the 20-40\% bin.

The MPC model has a number of mechanisms, including corona effects, energy
loss, and the ability to boost lower \pT partons to higher \pT (a unique
feature).  The calculation shown in Figure~\ref{fig:modelCompare} does a
better job of reproducing the overall shape of the \vTwo, though it is
systematically low.  It is important to note that this calculation is done for
one set of parameters, so it should be very interesting to see if the MPC can
better reproduce the data for a different set of parameters (the opacity of
the medium, for example).

The prevailing thought is that the high \pT behavior of the \vTwo is
due to energy loss mechanisms.  If this is true, the \RAA should be
sensitive to the geometry of the collision.  To test this behavior, we
seek to combine the two traditional geometric parameters (centrality,
or collision overlap, and angle of emission) into a single parameter,
a quantity which we will refer to as ``\leff''.  Details of the
calculation are described in~\cite{ref:leff}, as well as below.

\begin{figure}[t]
\begin{center}
\includegraphics*[width=0.45\textwidth]{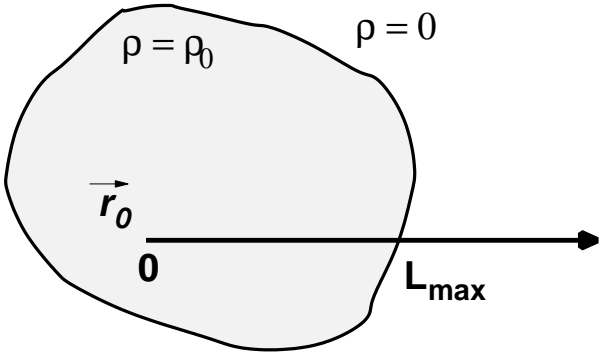}
\end{center}
\caption{Schematic of the calculation of $\int\leff$. See text for details.}
\label{fig:leff_calc}
\end{figure}

The Guylassy-Levai-Vitev (GLV) formalism can be used to calculate jet
energy loss for a set of scattering centers $\lbrace x_i \rbrace$,
where $x_i = (t_i, r_i)$.  In practice, an average over these centers
is performed.  As shown in Figure~\ref{fig:leff_calc}, if a static
uniform color charge density within some region ($\rho(x) = \rho_0$
and zero outside the region) is assumed, the resulting energy loss is
$\Delta E_{QCD} \propto \rho_0 L^2_{max}$.  More realistically, if the
density seen by the particle changes along the path, we have $\Delta
E_{QCD} \propto \int_0 \rho(L) L dL$ (which reduces to the quadradic
$L$ dependence for a constant density).  Application of this to a 1D
Bjorken expansion, with
\begin{equation}
\rho({\bf r},\tau) = \rho({\bf r},\tau_0) \frac{\tau_0}{\tau}
\end{equation}
and given a jet trajectory ${\bf r}(\tau) = {\bf r}_0 + {\bf v}(\tau-\tau_0)$ (assuming
$v \simeq c = 1$ for the jet), we have
\begin{equation}
L(\tau) = |{\bf r}(\tau) - {\bf r}_0| = \tau-\tau_0
\end{equation}
Therefore we have
\begin{eqnarray}
\Delta E_{QCD} & \propto & \int^{L_{max}}_0 \rho({\bf r}, \tau) L dL \\
& & \int^{L_{max}}_0 \frac{\rho({\bf r}, \tau_0) \tau_0 L}{\tau_0+L} dL 
\end{eqnarray}

This effective energy loss is calculated from the parton-density
weighted average of the length from hard-scatter\-ing origin to edge of
an ellipse.  Additionally, we perform a Glauber Monte Carlo sampling
of starting points to account for fluctuations in the location of the
hard-scattering origin of the particles' paths within the region of
overlap between the colliding nuclei.  The crucial feature of \leff is
that it is proportional to the energy loss sustained by the parton as
it traverses the medium.

\begin{figure*}[t]
\begin{center}
\includegraphics*[width=\textwidth]{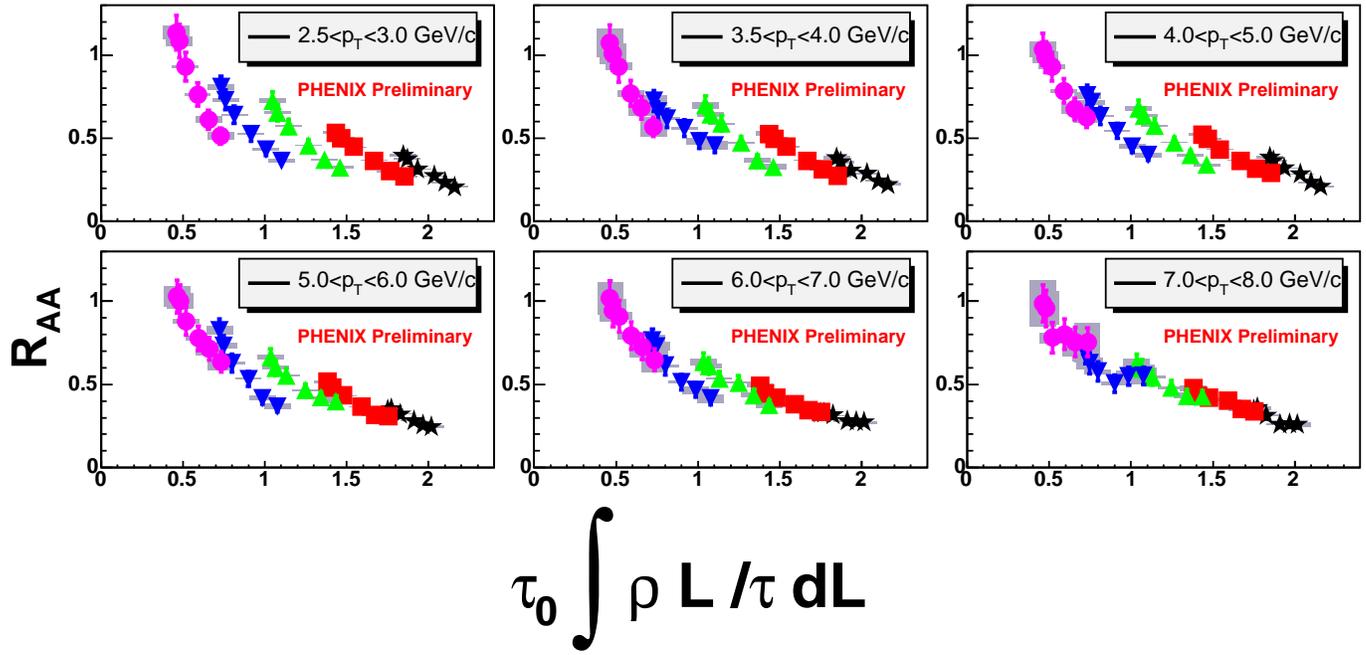}
\end{center}
\caption{$R_{AA}(\dphi,\pT)$ vs. $\leff$.  The panels correspond to different
  \pT ranges.  The solid circles are the most peripheral events, while the
  solid stars are the most central events.}
\label{fig:leff}
\end{figure*}

The resulting dependence of \RAA for all centralities and angles on
\leff is shown in Figure~\ref{fig:leff}.  If the observed \RAA arose
from only geometric effects, we would expect the data to exhibit a
universal dependence on \leff.  For low \pT, this is clearly not the
case; something more than just energy loss is taking place there.
However, when the \pT reaches 7 GeV/c and above, the \RAA data do
indeed appear to have a dependence on a single \leff curve.  This
apparent scaling strongly suggests that the dominant effect on \RAA at
high-\pT is energy loss.  These data can help to constrain energy loss
models, and perhaps help to understand the nature of that energy loss
(is it radiative, collisional, or some combination of both?).

\section{Conclusions}\label{concl}
We have presented the first measurement of high \pT \vTwo for \piz{}s.
It is now clear that the \vTwo at high \pT does decrease but to a
non-zero value.  Comparison of \vTwo with models suggest that the
dominant mechanism at work at high \pT is energy loss.  In addition,
we have presented the first measurement of \piz \RAA as a function of
angle with respect to the reaction plane.  When the \RAA data are
examined as a function of an effective path length through the medium,
the scaling that arises at high \pT also argues for energy loss as the
dominant mechanism at work.

%
%

\end{document}